# Ontology and Knowledge Management System on Epilepsy and Epileptic Seizures

Pedro Almeida, Paulo Gomes, Francisco Sales, Ana Nogueira, António Dourado

Departamento de Engenharia Informática, Faculdade de Ciências e Tecnologia, Pólo II, 3030-290, Coimbra, Portugal
pjga@student.dei.uc.pt, pgomes@dei.uc.pt, franciscosales@huc.min-saude.pt, ananog@student.dei.uc.pt, dourado@dei.uc.pt

**Abstract.** A Knowledge Management System developed for supporting creation, capture, storage and dissemination of information about Epilepsy and Epileptic Seizures is presented. We present an Ontology on Epilepsy and a Web-based prototype that together create the KMS.

**Keywords:** Semantic Web**,** Ontology, Knowledge Management System, EPILEPSIAE, Epilepsy

## 1  Introduction

The use of ontologies in the area of medical science has already a long way, and several Web references to medical ontologies for different areas of medicine can be found (UMLS [1], SNOMED [2], GALEN [3], etc.). Most of these ontologies are used in commercial knowledge management systems, not publicly accessed, and as a result, that knowledge cannot be shared with other ontologies and/or systems.

This project starts out the foundation for an open Ontology to represent knowledge about epilepsy and epileptic seizures and to build a system to store and manage the ontology so it can be evolved. Doctors, patients and researchers will have an easy access to complete and well-structured information on epilepsy and epileptic seizures and with their contributions may extend and build the most complete ontology on the subject. This work is part of the European project EPILEPSIAE [4], whose primary mission is to use information and communication technologies in order to enhance the safety of epileptic patients in their everyday life, allowing them to predict seizures and give them the ability to monitor their own risks.

## 2 Ontology

The developed base structure for the domain of epilepsy and epileptic seizures is based on the proposed diagnostic scheme for people with epileptic seizures and epilepsy by the International League Against Epilepsy (ILAE) [5]. This model was transformed into a formal model to be easily deployed through the Protégé [6] ontology editor tool, thereby creating the basis for the epilepsy ontology.

The key concepts that make up the epilepsy ontology are the following:

- **General concepts** - all common concepts of epilepsy and that can be used to describe other concepts.

- **Seizures Types** - all kinds of seizures types known to this day, in accordance with the proposed diagnostic scheme for people with epileptic seizures and epilepsy.

- **Epileptic Syndromes** – all known epileptic syndromes according to the proposed diagnostic scheme for people with epileptic seizures and epilepsy.

- **Electroencephalography** – all Electroencephalography related concepts.

The ontology has actually 145 concepts, 290 label annotations and 290 comments annotations. Annotations are in English and in Portuguese. Figure 1 shows one view of the epilepsy ontology.

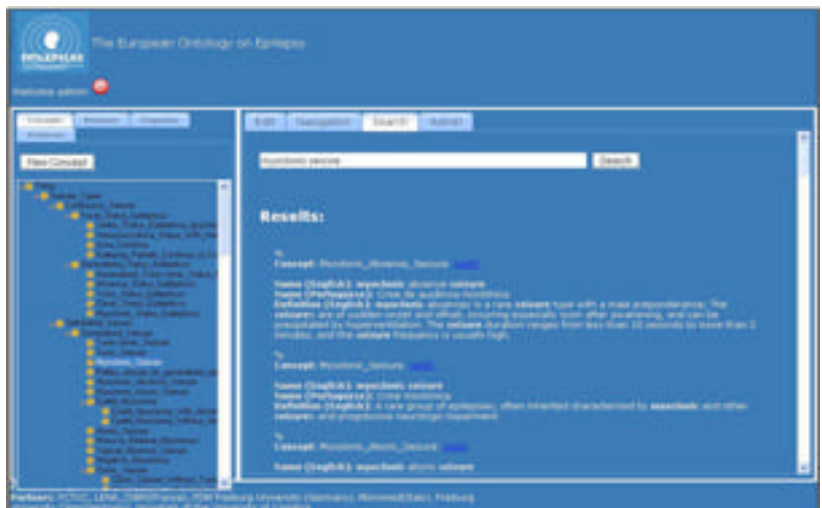

Figure 1. One view of the Epilepsy Ontology

## 3 Knowledge Management System

The main component in the system is the epilepsy ontology. The prototype developed intends to support the ontology from the point of view of storage, management and search.
The system is composed of five distinct modules: Management, Navigation Exploration, Search, EPILEPSIAE DB and Indexer. The Management, Navigation and Search modules are unified through a Web interface produced in JSP code,

Struts2 and JavaScript. The modular architecture schema of the system is presented in Fig. 2.

**Management module.** The management module is responsible for managing and to maintain the consistency of the ontology.

**Search module.** The search module is the module responsible for data query in the ontology. This module is responsible in encapsulating the SPARQL [7] code used for searches in ontologies.

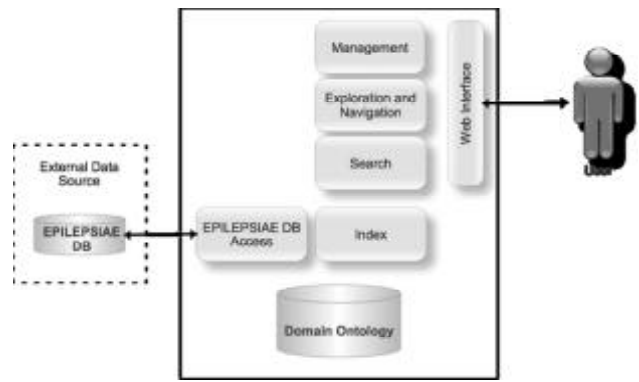

**Fig. 2.** Knowledge Management System Architecture

**Navigation module.** The Navigation and exploration module uses the Search module, to build a graphical image of the ontology in the form of a graph.

**EPILEPSIAE DB access module.** The EPILEPSIAE DB module access is responsible for the database connection to EPILEPSIAE and extracting text from certain fields present in some tables.

**Index module.** The index module makes use of the EPILEPSIAE DB access module described previous, in order to collect multiple resources, mostly in free text format, and index it.

**Web interface.** The Web interface as already mentioned, is the element that links the interaction through the Web browser and the management, navigation and search modules.

**Ontology Repository.** The Ontology repository is where the Epilepsy ontology is stored. It is composed of the free version of the AllegroGraph RDFStore [8] system, which provides a low-level API for managing and searching the ontology.

## 4 Conclusions and Future Work

The epilepsy ontology built as part of this project allows access to complete and well-structured information on epilepsy and epileptic seizures. The ontology can be at any time extracted from the AllegroGraph RDFStore [8] system in RDF [9] or OWL [10] and subsequently published on the Web. In this way, the ontology can be used by other users and systems based on the Semantic Web.

The prototype developed offers a complete platform for the management of the ontology, allowing to graphically navigate and explore the ontology structure. The search system allows to quickly finding ontology resources, with the possibility of suggestions if no results are found.

It is important to develop more features in relation to the management of ontologies : loading an ontology from an OWL [8] or RDF [9] file, to manage several ontologies at the same time, store detailed information about the changes that are being made to a given ontology, etc. The purpose of these new features is to create a full web-based ontologies management system for collaborative work.

Greater integration between the ontology/system and the EPILEPSIAE [4] database will be developed. Currently only the annotations present on the EPILEPSIAE database can be searched. A possible new application would use the ontology, in order to facilitate the introduction of information in the EPILEPSIAE database, i.e. suggesting words (concepts) to the user.

Because the area of epilepsy and epileptic seizures is constantly evolving, the ontology should reflect this evolution. It is important to continue communicating with medical doctors, so that new concepts are added to the ontology and their respective meaning, and most importantly, try to find new applications for the ontology, serving the patients, doctors and researchers.

The EPILEPSIAE project is developing algorithms and classifiers for seizure prediction, namely based on computational intelligence, such as neural networks and support vector machines, and their results are stored into the database. The ontological search of these classifiers and data will be developed, in order to support the development of seizure predictors for new patients.

**Acknowledgements.** This work is carried out in the EPILEPSIAE Project, EU FP7 Advanced ICT for Risk Assessment and Patient Safety, Grant 211713. The authors express their gratitude to the EU Commission for the financial support, namely for the scholarship of Pedro Almeida for his MsC thesis.